\documentclass[twocolumn,preprintnumbers,amsmath,amssymb]{revtex4}
\def\ket{\rangle}

\usepackage{graphicx}

\usepackage{amsmath}
\begin{document}
\preprint{}
\title{Deleting a marked item from an unsorted database with a single query}
\author{Yang Liu and Gui Lu Long}
\affiliation{$^1$ Key Laboratory of Atomic and Molecular NanoSciencs
and Department of Physics, Tsinghua University, Beijing 100084, P. R. China\\
$^2$Institute of Microelectronics, Tsinghua University, Beijing 100084, P. R. China\\
$^3$ Tsinghua National Laboratory for Information Science and
Technology, Beijing 100084, P. R. China}
\date{\today}

\begin{abstract}
In this Letter we present a quantum deletion algorithm that deletes
a marked state from an unsorted database of $N$ items with only a
single query. This algorithm achieves exponential speedup compared
with classical algorithm where $O(N)$ number of query is required.
General property of this deleting algorithm is also studied.
\end{abstract}
\pacs{ 03.67.-a} \maketitle

The merging of quantum mechanics and information theory has been
very fruitful in recent decades. A quantum computer can complete
difficult tasks that is hard for a classical computer. Remarkably,
Shor have shown that quantum computer can factorize a large integer
exponentially faster than a classical computer \cite{Shor}, and
Grover has proved that quantum computer can search an unsorted
database with a square-root speedup compared \cite{Grover1}.
Searching an unsorted database is a widely met scientific problem
\cite{longfoc}.
 There have been extensive studies on the generalization of the Grover algorithm
\cite{Grover2,Grover3, Boyer,Long,Phase1,Phase2,Phase3}.

A related task is to delete an item from a database. In a chain data
structure, if it is known that certain nodes does not satisfy the
data structure conditions or does not meet the algorithm demands,
then one should delete them from database for convenient searching
and visiting. For instance, in the PageRank algorithm \cite{LPage}
where the web pages were sorted according to their click amounts, we
usually select the former $N$ items which are clicked most
frequently from the magnanimity of information. Therefore we need to
add or delete the web pages with respect to the marked node in order
to dynamically preserve the former $N$ nodes. Then the former $N$
nodes form a heap structure which requires kinds of operations:
addition, deletion, search and visit. In some case, finding a marked
item from an unsorted database is prerequisite step for its next
disposal step. For example, if a database represents the components
of a giant machine, and we want to find out the bad component and
delete it from the database. Usually, deleting a marked item from an
unsorted database with $N$ items is equivalent to finding a marked
state from the database, and it requires $\emph{O}(N)$ steps in
classical compting.

In this Letter, we present a quantum algorithm that deletes an
marked item from  an unsorted database with only a single query.
Compared to its classical counterpart, it achieves an exponential
speedup. At first it may sounds very similar to quantum searching.
But, actually, it i very different. Indeed, here we do not require
the knowledge of the marked state: what counts is to delete it from
the database.

The abstract problem is: if there is an unsorted database with
$N=2^n$ items, one item satisfies a query function $f(\tau)=1$, and
all other items satisfy $f(x)=0$, the task is to delete the item
$\tau$ from the database.

In quantum mechanics, the problem becomes the following, in the
evenly superposed state $|\psi_0\ket$, which represents the database
consisting of all the items
\begin{eqnarray}
|\psi_0\rangle&=&\sqrt{\frac{1}{N}}\left(|0\rangle+|1\rangle+\cdots+|\tau\rangle+\cdots+|N-1\rangle\right
),\label{data}
\end{eqnarray}
one item $\tau$ satisfies the query $f(\tau)=1$, and all other items
satisfy $f(x)=0$. The task is to delete the item $\tau$ from the
superposed state (\ref{data}). Let $|c\rangle$ indicates a state
that is the sum over all $i$ which are not the marked state,
\begin{equation}
|c\rangle=\sqrt{\frac{1}{N}}\sum_{i\neq\tau}|i\rangle.
\end{equation}
In the two-dimensional spaces span by $|c\rangle$ and
$|\tau\rangle$, the initial state of the quantum computer
$|\psi_0\rangle$ may be re-expressed as follows
\begin{equation}
|\psi_0\rangle=\cos{\beta}|c\rangle+\sin{\beta}|\tau\rangle,
\end{equation}
where the coefficients
\begin{equation}
\sin\beta=\sqrt{\frac{1}{N}},\qquad\cos\beta=\sqrt{\frac{N-1}{N}}.\label{ephi}
\end{equation}

The quantum deletion algorithm consists of successive applications
of a quantum deletion subroutine, indicated as $S$ operation. The
$S$ operation contains four steps:

Step 1: Perform a conditional phase shift $e^{i\phi}$ to every
computational basis state except the marked state $|\tau\rangle$,
the action of this step is denoted as $I_{c}$
\begin{eqnarray}
I_c=I+(e^{i\phi}-1)\sum_{i\neq\tau}|i\rangle\langle i|.
\end{eqnarray}

Step 2: Perform the $n$-qubit Hadamard transform $W$, namely
$W=H^{\bigotimes n}$, where $H$ is the Hadamard transform on a
single qubit.

Step 3: Perform a conditional phase shift $e^{i\phi}$ to the
$|0\ket$ state and leaves all other basis states untouched. Denoting
this action as $I_0$, and it is
\begin{eqnarray}
I_0=I+(e^{i\phi}-1)|0\rangle\langle 0|.
\end{eqnarray}

Step 4: Perform the $n$-qubit Hadamard transform again.

The deletion operator $S$ can be represented in a matrix form in the
spaces span by $|c\rangle$ and $|\tau\rangle$,
\begin{eqnarray}
&&S=-WI_0WI_c\nonumber\\
&&=\left[\begin{array}{cc}-e^{i\phi}(1+(e^{i\phi}-1)\cos^2{\beta})&-(e^{i\phi}-1)\sin{\beta}\cos{\beta}\\
-e^{i\phi}(e^{i\phi}-1)\sin{\beta}\cos{\beta}&-e^{i\phi}+(e^{i\phi}-1)\cos^2{\beta}\end{array}\right].\nonumber\\
&&
\end{eqnarray}

In the above procedures, the two phases are equal, and this is
deeply rooted in the phase matching condition in quantum search
algorithm \cite{Phase1,Phase2,Phase3}. We have worked out the
explicitly the phase $\phi$, \begin{eqnarray}
\phi=2\arcsin\left(\frac{1}{2\cos\beta}\right). \end{eqnarray}

When applying $S$ to the database in Eq.(\ref{data}), through direct
calculation we obtain,
\begin{eqnarray}
|\psi_1\rangle
              &=&e^{i(\phi-\pi)/2}\;\sqrt{1\over
              N-1}\left(|0\ket+|1\ket+\cdots+|N-1\ket\right).
\end{eqnarray}
That is to say, except a global phase factor $(\phi-\pi)/2$, the
marked state $|\tau\rangle$ has been deleted from the unsorted
database with certainty using only one application of the deletion
operation $S$. The phase factor can be eliminated by applying a
phase rotation of $e^{(\pi-\phi)/2}$ to all the basis state. It is
striking that quantum deletion algorithm can delete a marked item
from an unsorted database with just a single query, which is
ultimate performance of any deletion algorithm. In comparison, a
classical computer requires $\emph{O}(N)$ queries to find the marked
state and then delete it from the database.


Repeating the deletion operation for a number of times, we find the
deletion algorithm will still be effective for certain times of
iteration. We investigate the periodic property of quantum deletion
algorithm and calculate the performance of $k$ times of $S$
iteration. Firstly, we can write out the expression of $S$ operator
in a diagonalized form
\begin{eqnarray}
S=U\Lambda U^\dag,
\end{eqnarray}
where $U$ is
\begin{eqnarray}
\sqrt{1\over R}\left[\begin{array}{cc}e^{-\frac{i\phi}{2}}(\cos\frac{\phi}{2}\cos\beta+\cos\beta^\prime)&-\sin\beta\\
\sin\beta&e^{\frac{i\phi}{2}}(\cos\frac{\phi}{2}\cos\beta+\cos\beta^\prime)\end{array}\right],\nonumber
\end{eqnarray}
\begin{eqnarray}
\Lambda&=&\left[\begin{array}{cc}-e^{i(\phi+2\beta^\prime)}&0\\0&-e^{i(\phi-2\beta^\prime)}\end{array}\right],\\
\beta^\prime &=&\arcsin(\sin\frac{\phi}{2}\cos\beta),\\
R&=&\sin^2\beta+(\cos\frac{\phi}{2}\cos\beta+\cos\beta^\prime).
\end{eqnarray}
Using Eq. (\ref{ephi}) and noting that
\begin{eqnarray}
&&\sin\frac{\phi}{2}=\frac{1}{2}\sqrt{\frac{N}{N-1}},
\hspace{10pt}\cos\frac{\phi}{2}=\frac{1}{2}\sqrt{\frac{3N-4}{N-1}},
\end{eqnarray}
we obtain
\begin{eqnarray}
\beta^{\prime}&=&\arcsin\frac{1}{2}=\frac{\pi}{6},\nonumber\\
R&=&\frac{3N+\sqrt{3N(3N-4)}}{2N}.
\end{eqnarray}
The $k$ successive operations of $S$ can be written analytically
\begin{equation}
S^k=U\Lambda^k
U^\dag=\left[\begin{array}{cc}s_{11}&s_{12}\\s_{21}&s_{22}\end{array}\right],
\end{equation}
where
\begin{eqnarray}
s_{11}&=&(-1)^ke^{ik\phi}(\cos\theta+i\sin\theta\sqrt{\frac{3N-4}{3N}}),\nonumber\\
s_{12}&=&(-1)^ke^{ik\phi}[\sin\theta\sqrt{\frac{1}{3(N-1)}}+i\sin\theta\sqrt{\frac{3N-4}{3N(N-1)}}],\nonumber\\
s_{21}&=&(-1)^ke^{ik\phi}[-\sin\theta\sqrt{\frac{1}{3(N-1)}}+i\sin\theta\sqrt{\frac{3N-4}{3N(N-1)}}],\nonumber\\
s_{22}&=&(-1)^ke^{ik\phi}(\cos\theta-i\sin\theta\sqrt{\frac{3N-4}{3N}}),
\end{eqnarray}
and $\theta=2k\beta^\prime=\frac{k\pi}{3}$. The final analytic
expression for $S^k$ is
\begin{widetext}
\begin{eqnarray}
S^k=(-1)^ke^{ik\phi}\left[\begin{array}{cc}\cos\theta+i\sin\theta\sqrt{\frac{3N-4}{3N}}
&
\sin\theta\sqrt{\frac{1}{3(N-1)}}+i\sin\theta\sqrt{\frac{3N-4}{3N(N-1)}} \\
-\sin\theta\sqrt{\frac{1}{3(N-1)}}+i\sin\theta\sqrt{\frac{3N-4}{3N(N-1)}}
& \cos\theta-i\sin\theta\sqrt{\frac{3N-4}{3N}}\end{array}\right].
\label{eqn1}
\end{eqnarray}
\end{widetext}
As shown in Table. \ref{table1},  functions $\sin\theta$ and
$\cos\theta$ vary periodically with a period of 6 in $k$, and
functions of $(-1)^k\sin\theta$ and $(-1)^k\cos\theta$ vary
periodically with a period 3. Hence $S^k$ is a periodic function of
$k$ with 3 as its period. We now look at the three different
situations.
\begin{table*}
\caption{periodic property of some trigonometric functions }
\begin{ruledtabular}
\begin{tabular}{c|cccccccccccc}
k&1&2&3&4&5&6&7&8&9&10&11&12\\
\hline $\sin\theta$&$\frac{\sqrt{3}}{2}$&$\frac{\sqrt{3}}{2}$&$0$&$-\frac{\sqrt{3}}{2}$&$-\frac{\sqrt{3}}{2}$&$0$&$\frac{\sqrt{3}}{2}$&$\frac{\sqrt{3}}{2}$&$0$&$-\frac{\sqrt{3}}{2}$&$-\frac{\sqrt{3}}{2}$&$0$\\
$\cos\theta$&$\frac{1}{2}$&$-\frac{1}{2}$&$-1$&$-\frac{1}{2}$&$\frac{1}{2}$&$1$&$\frac{1}{2}$&$-\frac{1}{2}$&$-1$&$-\frac{1}{2}$&$\frac{1}{2}$&$1$\\
\hline
$(-1)^k\sin\theta$&$-\frac{\sqrt{3}}{2}$&$\frac{\sqrt{3}}{2}$&$0$&$-\frac{\sqrt{3}}{2}$&$\frac{\sqrt{3}}{2}$&$0$&$-\frac{\sqrt{3}}{2}$&$\frac{\sqrt{3}}{2}$&$0$&$-\frac{\sqrt{3}}{2}$&$\frac{\sqrt{3}}{2}$&$0$\\
$(-1)^k\cos\theta$&$-\frac{1}{2}$&$-\frac{1}{2}$&$1$&$-\frac{1}{2}$&$-\frac{1}{2}$&$1$&$-\frac{1}{2}$&$-\frac{1}{2}$&$1$&$-\frac{1}{2}$&$-\frac{1}{2}$&$1$\\
\end{tabular}
\end{ruledtabular}\label{table1}
\end{table*}

Case 1. When $k=3m+1$, i.e. $k=1,4,7$, $10,13\cdots$, Eq.
(\ref{eqn1}) reduces to the following form
\begin{small}
\begin{equation}
S^k=e^{ik\phi}\left[\begin{array}{cc}-\frac{1}{2}-\frac{i}{2}\sqrt{\frac{3N-4}{N}}&-\frac{1}{2}\sqrt{\frac{1}{N-1}}-\frac{i}{2}\sqrt{\frac{3N-4}{N(N-1)}}\\
\frac{1}{2}\sqrt{\frac{1}{N-1}}-\frac{i}{2}\sqrt{\frac{3N-4}{N(N-1)}}&-\frac{1}{2}+\frac{i}{2}\sqrt{\frac{3N-4}{N}}\end{array}\right].
\end{equation}
\end{small}
After $S^k$, the state becomes
\begin{eqnarray}
|\psi_k\rangle&=&S^k(\cos\beta|c\ket+\sin\beta|\tau\ket)\\
&=&e^{i[(k-1/2)\phi-\pi/2]}\sqrt{1\over
N-1}(|0\ket+|1\ket+\cdots+|N-1\ket).\nonumber
\end{eqnarray}

Hence this type of iteration process accomplishes the following
transformation eventually
\begin{equation}
\sqrt{\frac{1}{N}}\sum^{N-1}_{i=0}|i\rangle\xrightarrow{S^{3m+1}}e^{i[(k-1/2)\phi-\pi/2]}\sqrt{\frac{1}{N-1}}\sum_{i\neq\tau}|i\rangle.
\end{equation}
We can see that if the number of iteration is $k=3m+1$, where $m$ is
any nonnegative integer,  after $k$ deletion iterations the marked
state $|\tau\rangle$ will be successfully deleted from the unsorted
database. The global phase factor can be left alone, or by a
simultaneous phase rotation of $e^{-i[(k-1/2)\phi-\pi/2]}$ to all
basis states to eliminate the total phase.

Case 2. When $k=3m+2$, i.e. $k=2,5,8,11,14\cdots$, Eq. (\ref{eqn1})
can be rewritten as
\begin{small}
\begin{equation}
S^k=e^{ik\phi}\left[\begin{array}{cc}-\frac{1}{2}+\frac{i}{2}\sqrt{\frac{3N-4}{N}}&\frac{1}{2}\sqrt{\frac{1}{N-1}}+\frac{i}{2}\sqrt{\frac{3N-4}{N(N-1)}}\\
-\frac{1}{2}\sqrt{\frac{1}{N-1}}+\frac{i}{2}\sqrt{\frac{3N-4}{N(N-1)}}&-\frac{1}{2}-\frac{i}{2}\sqrt{\frac{3N-4}{N}}\end{array}\right].
\end{equation}
\end{small}
After $k$ times of $S$ iteration, the initial evenly distributed
state becomes
\begin{eqnarray}
|\psi_k\rangle&=&S^k\left[\begin{array}{c}\cos\beta\\\sin\beta\end{array}\right]
=e^{ik\phi}\left[\begin{array}{c}e^{i(\pi+\phi)}\sqrt{\frac{N-1}{N}}
\\  e^{i\pi}\sqrt{\frac{1}{N}}\end{array}\right],
\end{eqnarray}
where
\begin{eqnarray}
e^{i(\pi+\phi)}&=&-\frac{N-2}{2(N-1)}+\frac{i\sqrt{N(3N-4)}}{2(N-1)}.
\end{eqnarray}
This type of transformation can be represented consequently as
\begin{equation}
\sqrt{\frac{1}{N}}\sum^{N-1}_{i=0}|i\rangle\xrightarrow{S^{3m+2}}
{e^{i[\pi+(k+1)\phi]}\over
\sqrt{N}}\sum_{i\neq\tau}|i\rangle+{e^{i(\pi+k\phi)}\over
\sqrt{N}}|\tau\rangle.
\end{equation}
It implies that after $k=3m+2$ number of iterations the marked state
can not be deleted except for a different phase change of the marked
state relative to other states.

Case 3. When $k=3m+3$, i.e. $k=3,6,9$, $12,15\cdots$, we may reduce
Eq. (\ref{eqn1}) to
\begin{equation}
S^k=e^{ik\phi}\left[\begin{array}{cc}1&0\\0&1\end{array}\right]=e^{ik\phi}I,
\end{equation}
so except a global phase factor $e^{ik\phi}$, $k$ times of iteration
$S$ leaves the state of the system state  in the evenly distributed
state.
\begin{equation}
|\psi_k\rangle=S^k\left[\begin{array}{c}\cos\beta\\\sin\beta\end{array}\right]
=e^{ik\phi}\left[\begin{array}{c}\sqrt{\frac{N-1}{N}}\\\sqrt{\frac{1}{N}}\end{array}\right].
\end{equation}
This type of iteration process can be expressed as
\begin{equation}
\sqrt{\frac{1}{N}}\sum^{N-1}_{i=0}|i\rangle\xrightarrow{S^{3m+3}}e^{ik\phi}{1
\over \sqrt{N}}\sum^{N-1}_{i=0}|i\rangle.
\end{equation}
We can see if the number of iteration is $k=3m+3$,  the initial
evenly distributed state of the unsorted database would remain.

\begin{figure}[here]
\begin{center}
\includegraphics[width=7.5cm]{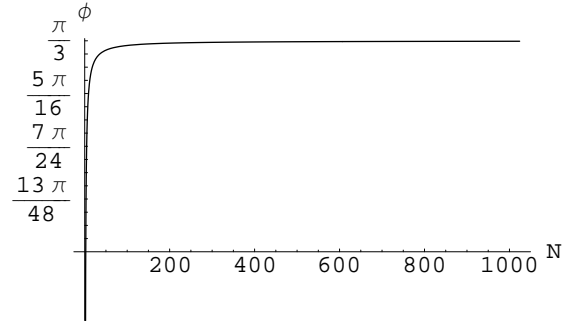}
\caption{The phase angle $\phi$ versus the datasize $N$.}\label{f1}
\end{center}
\end{figure}

Next we present an approximate deletion algorithm that uses a fixed
phase rotation. In the unsorted database search problem, the
Grover's algorithm which finds a marked item with high probability,
whereas the Long algorithm finds the marked item with certainty by
using datasize dependent phase rotations \cite{Long}. A schematic
plot for the phase angle $\phi$ versus database size is given in
Fig. \ref{f1}.
The approximate deletion algorithm replaces the datasize phase angle
with a constant phase angle, namely
\begin{eqnarray}
\phi_0=\lim_{N\rightarrow \infty}\phi={\pi\over3},
\end{eqnarray}
that is, steps 1 and 3, the phase rotations are replaced by phase
rotations through $\pi/3$. Then the $S$ matrix becomes
\begin{eqnarray}
S=\left[\begin{array}{cc}\frac{N-2}{2N}-i\frac{\sqrt{3}}{2}&\frac{\sqrt{N-1}}{2N}-i\frac{\sqrt{3(N-1)}}{2N}\\
\frac{\sqrt{N-1}}{N}&\frac{-2N+1}{2N}-i\frac{\sqrt{3}}{2N}\end{array}\right].
\end{eqnarray}
After the operation of $S$, the norm of the component of the marked
state becomes $N^{-3/2}$
\begin{eqnarray}
\lim_{N\rightarrow \infty}\frac{1}{N^{3/2}}=0.
\end{eqnarray}
Therefore under the large datasize limit, the marked state component
approaches zero, or say, the marked state is near completely
deleted.

In summary, a quantum deletion algorithm with certainty is present.
This algorithm deletes the marked state from an unsorted database.
We have shown that the quantum deletion algorithm completes deletion
with only a single query, in contrast to the $\emph{O}(N)$ steps
required in classical computing. Further it was discovered that the
deletion operation is periodic and has period of 3. Moreover, we
have present an approximate quantum deletion algorithm in which the
phase rotation has a fixed value of $\pi/3$.

{\bf Acknowledgement}

 This Work is supported by the National Natural Science Foundation of
China (Grant Nos. 10325521), the national Basic Research Program of
China (2006CB921106) , the Specialized Research Fund for the
Doctoral Program of Education Ministry of China(20060003048) .

\end{document}